# Neurological Consequences of COVID-19 Infection

Jiabin Tang[1,2], Shivani Patel[1], Steve Gentleman[1], Paul M. Matthews[1,2]

1 Department of Brain Sciences, Faculty of Medicine, Imperial College London, London, UK

2 UK Dementia Research Institute, Imperial College London, London, UK

**To whom correspondence should be addressed:**

Paul M. Matthews, MD PhD

E502, 5th Floor, Burlington Danes, Hammersmith Campus, Department of Brain Sciences, Faculty of Medicine, Imperial College London, London, UK

Tel: 0207 594 2855; Email: p.matthews@imperial.ac.uk

**Abstract**

COVID-19 infections have well described systemic manifestations, especially respiratory problems. There are currently no specific treatments or vaccines against the current strain. With higher case numbers, a range of neurological symptoms are becoming apparent. The mechanisms responsible for these are not well defined, other than those related to hypoxia and microthrombi. We speculate that sustained systemic immune activation seen with SARS-CoV-2 may also cause secondary autoimmune activation in the CNS. Patients with chronic neurological diseases may be at higher risk because of chronic secondary respiratory disease and potentially poor nutritional status. Here, we review the impact of COVID-19 on people with chronic neurological diseases and potential mechanisms. We believe special attention to protecting people with neurodegenerative disease is warranted. We are concerned about a possible delayed "pandemic" in the form of an increased burden of neurodegenerative disease after acceleration of pathology by systemic COVID-19 infections.

# Introduction

Over the past few months, the world has been transformed by the highly infectious coronavirus disease 2019 (COVID-19) - the third recorded global coronavirus (CoV) outbreak. Initially reported in Wuhan, China, COVID-19 has rapidly spread across the entire world.[1,2] Previously, the related severe acute respiratory syndrome (SARS) virus began in China in 2003, followed by Middle East respiratory syndrome (MERS) that started in Saudi Arabia in 2012.[1] SARS-CoV and MERS-CoV originated from bats that infected civet cats and camels, respectively, with human transmission thereafter. Likewise, SARS-CoV-2, responsible for COVID-19, is also thought to originate from bats, although the intermediate host has yet to be determined. Snakes or pangolins may be potential intermediate hosts.[3,4]

On February 11, 2020, WHO classified the current virus as SARS-CoV-2,[5] due to its high homology to SARS-CoV, and the associated disease was named COVID-19.[6] Three major risk factors have been identified: sex (male), age (≥60 years), and underlying health conditions (especially pneumonia, asthma, diabetes and heart disease).[3,7] COVID-19 has a lower severity and mortality rate than SARS, but it is much more contagious.[3]

CNS disorders that are secondary to infection by viral agents (e.g., encephalitis lethargica) are well described.[8] In some cases, direct viral infection of cells in the CNS occurs, e.g. herpes simplex virus-1 (HSV-1) causing herpes simplex encephalitis (HSE).[9] Currently, it is unknown whether there are long-term neurological complications with COVID-19. Nonetheless, such incidences in past pandemics should be considered to promote awareness of possible unexpected SARS-CoV-2 consequences in the CNS.

# Coronaviruses – An Old Enemy

Coronaviruses (CoVs) are large enveloped positive-stranded RNA viruses that induce enteric and respiratory diseases in animals and humans.[10] There are 6 CoV members that cause human disease: SARS-CoV, MERS-CoV, HCoV-HKU1, HCoV-NL63, HCoV-OC43, and HCoV-229E.[6] CoVs normally infect the upper respiratory tract in humans, with one strain causing symptoms of the common cold. They can also infect the lower respiratory tract in susceptible individuals, such as newborns, infants, the elderly and those who are immunocompromised or with underlying health conditions, leading to pneumonia, asthma exacerbation, respiratory distress syndrome, SARS or MERS.[11,12]

CoVs can be divided into four groups, or genera: (1) alpha-CoVs, (2) beta-CoVs, (3) gamma-CoVs, and (4) delta-CoVs. Alpha- and beta-CoVs infect humans.[13,14] Like its SARS-CoV homolog, SARS-CoV-2 is also a member of the beta-CoV genus.[3]

Replication and survival of the virus, is dependent on its ability to bind and 'hijack' host cell machinery, and mimics the behaviour of SARS-CoV and MERS-CoV.[15] SARS-CoV-2 enters cells by binding to angiotensin converting enzyme 2 (ACE2) receptor, which is highly expressed in the epithelial cells of respiratory and gastrointestinal tract, as well as heart, kidney and bladder.[16-18] Infection is initiated when the envelope spike glycoprotein on the SARS-CoV-2 plasma membrane binds to host cellular receptors, ACE2, which leads to fusion of membranes and entry of the virion into the host cell via endocytosis. The viral RNA genome is then released into the cytoplasm, enabling replication of the viral genome and synthesis of glycoproteins and nucleocapsid proteins for the envelope. The latter components are packaged into virion-containing vesicles, which are released subsequently into the extracellular environment via exocytosis, enabling infection propagation in the host (Figure 1).[19]

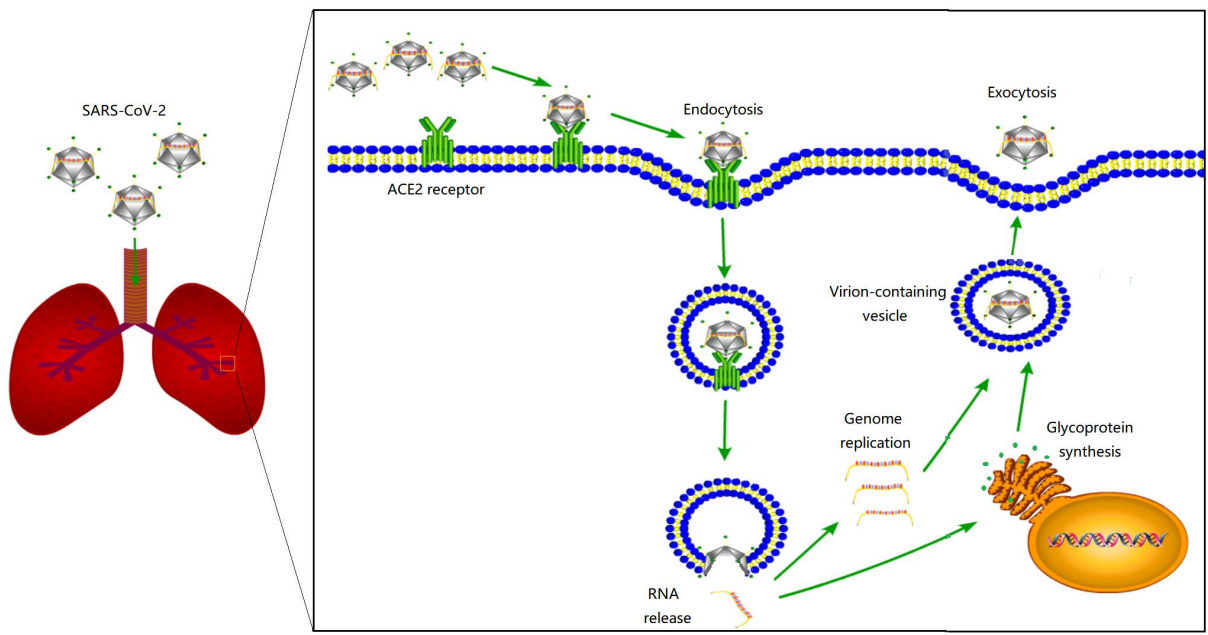

**Figure 1. Potential underlying mechanisms of SARS-CoV-2 'hijacking' host cell machinery.** SARS-CoV-2 binds to ACE2 receptor and enters the cell via endocytosis. Then, the viral RNAs are released to cytoplasm, undergoing replication and promoting glycoprotein synthesis. Finally, the new SARS-CoV-2 virus will be released to the extracellular environment by exocytosis. ACE2, angiotensin converting enzyme 2.

## Potential Consequences of CoV Infection for the Brain

Due to misdirected host immune responses, respiratory viruses can harm the CNS via autoimmunity.[20] Human CoVs have a similar molecular structure and mode of replication as animal CoVs, such as porcine hemagglutinating encephalitis virus

(PHEV), feline coronavirus (FCoV), and mouse hepatitis virus (MHV).[21-23] CoVs thus are well recognized as being able to invade the CNS and cause viral encephalitis, infectious toxic encephalopathy or acute cerebrovascular disease.[24] [25] A CoV-induced encephalomyelitis mouse model was established several years ago, indicating that CoV is able to cause progressive demyelinating neurodegenerative disease.[26] Another study with mouse models has also shown persistence of HCoV-OC43 RNA even one year post-infection in the CNS after intracerebral inoculation. The mice had abnormal reflexes and reduced activity in an open field test, as well as atrophy and neuronal loss in the hippocampus.[27] HCoV-229E and HCoV-OC43,[28-30] as well as SARS-CoV[31] [32] are all neuroinvasive and neurotropic in humans. This has been confirmed by the presence of coronaviral RNA in the human brain, demonstrating that these respiratory pathogens contribute to neurovirulence.[30] [33]

On 4th March, 2020, gene sequencing confirmed the presence of SARS-CoV-2 in the cerebrospinal fluid (CSF) of a 56-year-old patient in Beijing Ditan Hospital. The patient was diagnosed with viral encephalitis, suggesting that SARS-CoV-2 could potentially invade the CNS, rather than just acting secondary to a systemic immune response.[34] This was the first piece of evidence for direct nervous system invasion by SARS-CoV-2. Subsequently, SARS-CoV-2 has also been observed in neural and capillary endothelial cells in frontal lobe tissue at post-mortem examination.[35]

Neurological impairments resulting from SARS-CoV-2 infection include anosmia, ageusia, encephalopathy, prominent agitation, confusion and corticospinal tract signs, as reported in a study where 84% (49/58) of patients experienced these abnormalities.[36] [37] Severe infection has also resulted in stroke in some patients.[38] [39] Evidence is building that SARS-CoV-2 can enter the brain through (1) hematogenous dissemination - secondary inflammatory activation or ischemia secondary to a systemic coagulopathy, (2) neuronal retrograde dissemination - neuronal infection in the periphery and subsequent use of neural transport machinery (Figure 2).[2] [40]

Potential routes of hematogenous dissemination of SARS-CoV-2 may start with virus entry into the human airways and trans-epithelial entry of the virus into the bloodstream. Monocytes can become activated as a direct virus response and release factors such as MMP9 to increase blood brain barrier (BBB) permeability, as well as pro-inflammatory factors such as TNF-α to upregulate ICAM-1 expression on endothelial cells. This eases the passage of infected and activated monocytes into the CNS,[41] and SARS-CoV-2 is then able to pass through the loosened BBB.[6] Infiltrated infected monocytes activate microglia that produce chemokines (CCL5, CXCL10, CXCL11) and induce chemoattraction of activated T cells or other monocytes. In the meantime, astrocytes may produce other chemokines (CCL2, CCL5 and CXCL12) to aid the recruitment of more infected leukocytes, ultimately initiating a neuroinflammatory loop to cause immune-mediated neuropathology and a systemic inflammatory storm.[11] Hence, systemic inflammation with prolonged hypoxia induces persistent and uncontrolled neuroinflammation, which may underlie clinical manifestations present in patients recovering from pneumonia, including cognitive impairment, behavioural changes and delirium due to cortical deficits,[6] as well as deficits in attention and memory due to hippocampal damage.[6 42]

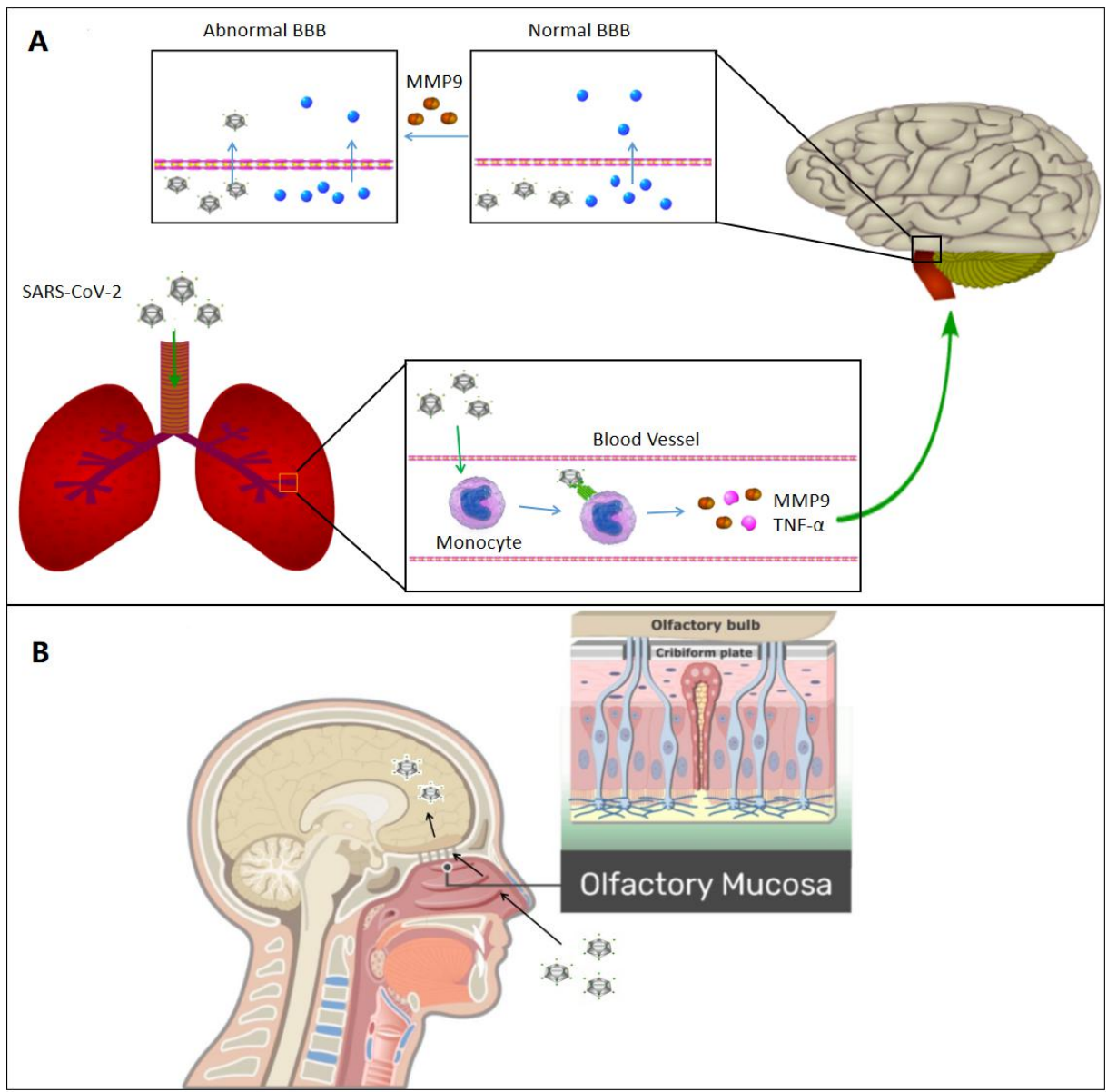

**Figure 2. Potential entry routes of SARS-CoV-2 invading the brain. (A)** Via hematogenous dissemination. SARS-CoV-2 could activate monocytes, increasing MMP9 and TNF-α level, and therefore increase the permeability of blood brain barrier (BBB). **(B)** Via neuronal retrograde dissemination. SARS-CoV-2 can go through olfactory mucosa and enter the brain directly via olfactory bulb and olfactory nerve. (B) is reproduced from [43].

For the neuronal retrograde dissemination method, nasal entry allows viral migration through the neuroepithelium of the olfactory mucosa to the mitral cells, and olfactory bulb thereafter.[44] Subsequently, SARS-CoV-2 may disseminate to the hippocampus and other brain regions via olfactory nerves[11,45] or via synaptic transmission.[46] SARS-CoV-2 is also likely to affect the brainstem heavily[47,48] in a similar manner to SARS-CoV and MERS-CoV,[49] since neural and glial ACE2 expression is prevalent in the brainstem, especially in areas that regulate cardiovascular function.[50,51] Many studies have claimed that the SARS-CoVs seems to have neurotoxicity effects with the interaction of ACE2 in the medullary brainstem,[52] and could induce neurodegeneration, neuroinflammation and astrogliosis.[53,54] The presence of anosmia and ageusia in SARS-CoV-2 infected subjects may also support the hypothesis of nasal route entry of the virus.[37,55] However, Bostanciklioğlu M. argued against it, claiming the glial-lymphatic pathway containing olfactory lymphatic vessels as a direct entry route for SARS-CoV-2 to the brain, which is intriguing and worthwhile for further investigation.[56]

Some scientists are aiming to inhibit ACE2 interaction to treat COVID-19, and ACE2-transgenic mice as well as non-human primates are being developed.[57] However, studies have shown that lung injury can be worsened upon decreased ACE2 expression levels,[58] and the overexpression of ACE2 in neuron cells could protect the brain from ischemic stroke,[59,60] so ACE2 inhibitors should be used with caution. A recently published study using human recombinant soluble ACE2 turned out to inhibit SARS-CoV-2 infection successfully in both human blood vessel organoids and human kidney organoids,[61] which could be a promising therapeutic direction.

## COVID-19 in Patients with Neurodegenerative Diseases

Greater number of co-morbidities and the increased neurodegenerative prevalence with age may mean that patients with neurodegenerative diseases are at greater risk of COVID-19. Patients with advanced Parkinson's Disease (PD) could be particularly vulnerable to SARS-CoV-2, due to rigidity of respiratory muscles, impairment of cough reflex and pre-existing dyspnea.[62] Optimal neurological care has also been compromised due to the pandemic and strain on health care systems.[63] Previously, antibodies against CoVs have been detected in the CSF of PD patients,[64] suggesting that virions could penetrate the blood brain barrier and potentially contribute to the progression of PD.[62] Dopaminergic neurons abundantly express ACE2 on their membranes. Hence, infiltration of SARS-CoV-2 may exacerbate PD-related symptoms.

In addition, with more than 50 million people having dementia, a risk of double hit of dementia and COVID-19 could further exacerbate the quality of life for patients.[65] Currently, there is limited data of COVID-19 patients with dementia, but these patients have been reported to be at increased risk for COVID-19 compared to patients without dementia.[66] There are two possible mechanisms that could initiate potential AD development after SARS-CoV-2 infection. Firstly, viral infection into CNS could directly cause neuroinflammation, and theoretically induce microglia and astrocyte activation, as well as disruption of proteostasis (amyloid production and tau phosphorylation).[67,68] Secondly, infiltration of various peripheral cytokines (IL-10, IL-1β, IFN-γ, TNF-α) into the CNS after SARS-CoV-2 incubation also may lead to neuroinflammation and even encephalitis (Figure 3).[69-71] Numerous viruses have been reported to be closely related to the risk of AD within cohort studies, including herpes simplex virus type 1 (HSV1), human herpesvirus 6 (HHV6) and hepatitis C virus (HCV), although there is still evidence disputing this.[68,72,73] According to incomplete retrospective statistics, 5% to 6% severe COVID-19 patients have been described to have cerebrovascular complications, 27% with brain insults, and 36% - 91% with neurologic disorders.[38,74-76] Besides, discoveries have also shown probable relationship between AD genes and increased COVID-19 mortality, and secondary effects of other organ system failure or side-effects of sedatives could also put the patients at exacerbated risk for AD.[77,78]

During the SARS and MERS pandemics, evidence has shown the neurotropism of CoV,[53,54,79] and the viral infection has been reported to contribute to the pathophysiology of brain disorders like encephalopathy, dementia, neuropathy and psychosis.[80,81] Because of the nature of chronic diseases, the symptoms may appear years after patients have recovered from viral infection, which makes it easy to overlook the connection. Patients with diabetes, which is one of the major comorbidities for COVID-19, appear to predispose to AD as well according to genetic etiologies.[82] In COVID-19 patients, various CNS symptoms have already been reported, such as headache, consciousness alteration and epilepsy, while brain imaging also showed ischemic infarct, hemorrhage and acute encephalopathy.[83,84] Moreover, the rate of CNS disorders is concordant with the respiratory problem status.[38] Interestingly, memantine, a common drug for AD, appears to have antiviral potential against SARS-CoV-2,[85] indicating the connection of underlying mechanisms between COVID-19 and AD. Memantine, an N-Methyl-D-aspartic acid (NMDA) receptor inhibitor, appears to be a good option for clinical studies, since it could inhibit glutamate release, which is a putative neurotoxic effect of depleted ACE2 (Figure 4).[85,86] Memantine has already been reported to be effective against CoV in 2004.[87] In addition, corticosteroid administration has shown progressive clinical improvement in a SARS-CoV-2 infected encephalitis patient, suggesting another possible therapeutic direction.[88]

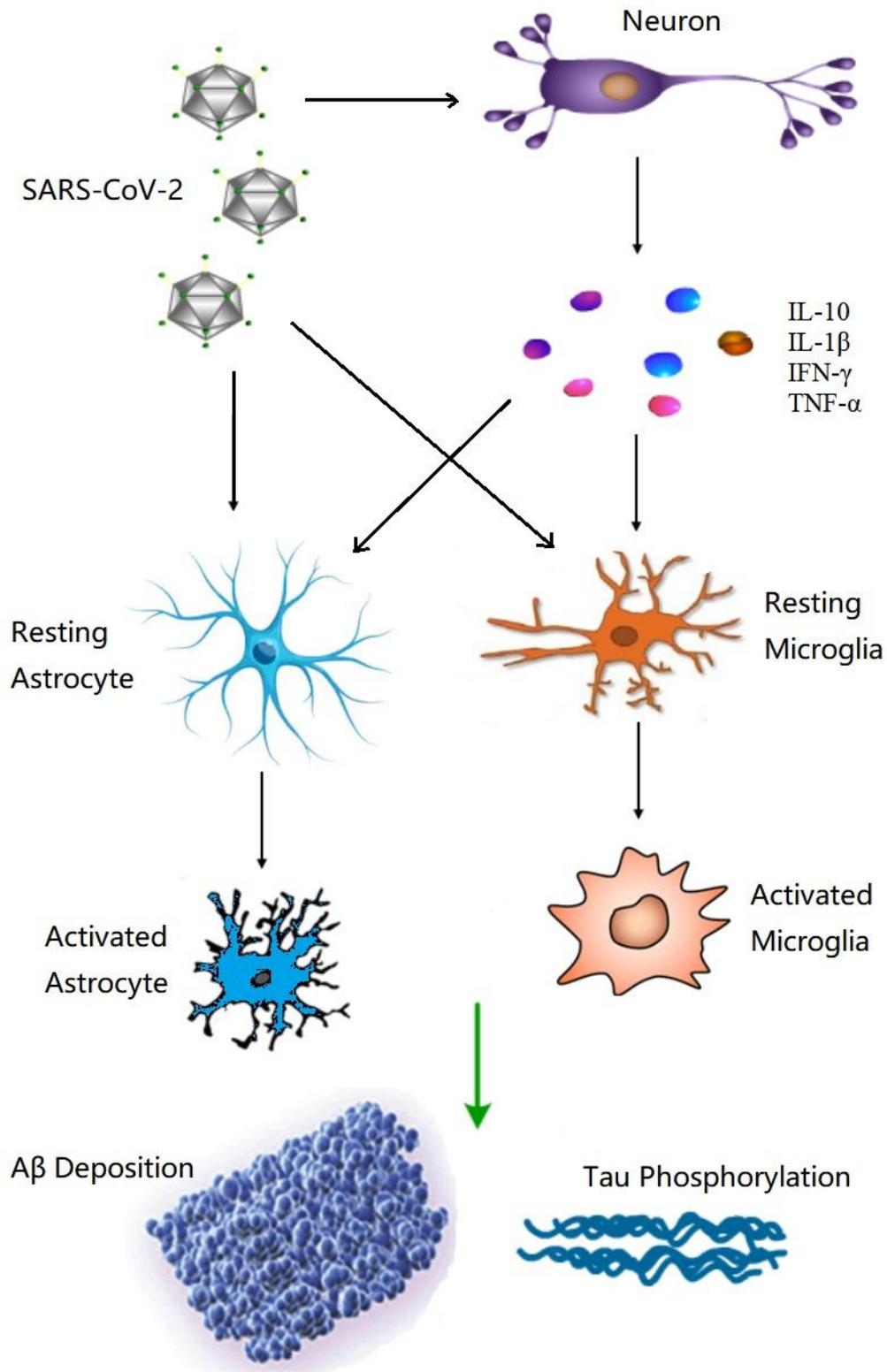

**Figure 3. Potential underlying mechanisms that SARS-CoV-2 initiates AD development.** SARS-CoV-2 could induce the activation of astrocytes and microglia directly, and could also promote the release or infiltration of peripheral cytokines (IL-10, IL-1β, IFN-γ, TNF-α), therefore leading to glial activation. CNS homeostasis could then be disrupted, accelerating Aβ deposition and tau phosphorylation, which are the hallmarks of AD progression.

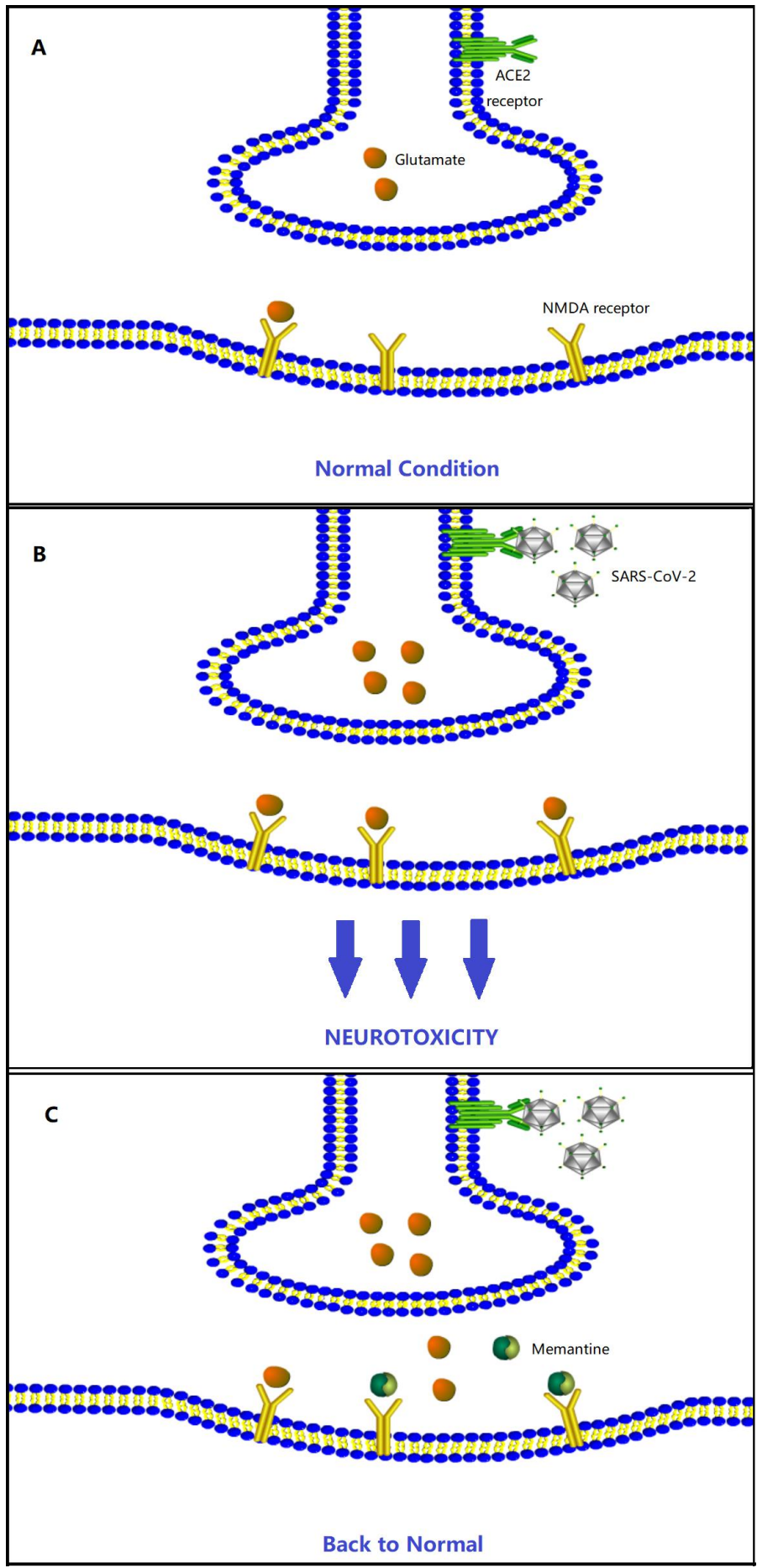

**Figure 4. Potential underlying mechanisms of memantine against SARS-CoV-2. (A)** In normal condition, ACE2 is supposed to function as an inhibitor for glutamate release, preventing glutamate over-release in healthy people. **(B)** SARS-CoV-2 could bind to ACE2 receptor and inhibit ACE2 function, leading to glutamate over-release. Glutamate will then interact with NMDA receptor and cause neurotoxicity. **(C)** Memantine could bind to NMDA receptors, acting as an antagonist to glutamate. NMDA, N-Methyl-D-aspartic acid. ACE2, angiotensin converting enzyme 2.

The SARS-CoV-2 infection could disrupt the brain homeostasis, making the CNS more vulnerable for AD progression even if asymptomatic for decades. Apart from this, interrupted medication of current AD patients could augment neurorehabilitation needs as well. Nevertheless, in a transgenic mouse model, apolipoprotein D, typically overexpressed in PD and AD patients, showed resistance effects against CoV-induced neurodegeneration, which makes us think whether the neurological disorders could protect the individual against CoV in a certain degree.[89][90]

## COVID-19 VS Neurodegenerative Diseases - Dilemmas and Directions

The COVID-19 pandemic has been unprecedented in the modern age because of its high infection rate, extent and range of morbidities and, in those more vulnerable, mortality. The need for rapid, more efficient conducted research has led to new way of working through national and international collaborative mechanistic studies and efforts to develop therapies for management and effective vaccines. Increasing the speed of disseminating research data has been fundamental to this. Clinicians, as well as scientists, should be aware and participate in the development of platforms such as the UK National Institute for Health Research National Bioresource, which shares and deploys the data worldwide for neurological disorders associated with COVID-19.[91]

Healthcare systems have been forced to modify medical care priorities due to the COVID-19 pandemic in ways that have had a dramatic impact on patients with chronic neurodegenerative conditions.[92] Telemedicine has become a preferred option to deliver urgent and ongoing healthcare.[93][94] However, mental wellbeing for patients with neurodegenerative disease is at risk from the effect of self-isolation and social distancing themselves.[94][95] Physical wellbeing and resilience may be impacted negatively by reductions in physical activity and social interactions; mere lacking regular walking and social interactions, as well as access to physiotherapists or fitness classes has consequences.[96] In addition, many patients and caregivers have not been willing to go to hospital since the onset of pandemic, because of the fear of in-hospital infection; data showed an approximately 50% reduction of 'non-essential' brain disease admissions.[97][98] The normal aging process could also be exacerbated by SARS-CoV-2 infection, in turn leading to higher risk of PD or AD.[99] New approaches for patient management that can accommodate changing clinical practices need to be accelerated. Emerging software tools directed to both neurologists and patients that support the management of chronic diseases and contribute to better mental health are amongst these.[74,100]

To date, there is no specific antiviral treatment approved for COVID-19 and no vaccination is currently available. Most hospitals are still relying on symptomatic treatment according to the doctors' experience, or treatment for comorbidities. Some nonspecific therapies have been proposed based on past experience with SARS and MERS. Clinical tests have been initiated to study the potential of existing antiviral agents,[101,102] interferons,[102,103] immunotherapies,[104] as well as some Chinese medicines.[105,106] Inhibitors of viral RNA polymerase include remdesivir and favipiravir, as well as inhibitors of viral protein synthesis and maturation include lopinavir and ritonavir.[107,108] Promising emerging medicines include a CRISPR-Cas13-based strategy - prophylactic antiviral CRISPR in human cells (PAC-MAN) - has been reported to degrade RNA from SARS-CoV-2 and inhibit viral replication.[109] It is significant to discover if the neurological symptoms are caused by viral entry or secondary immune response, since the treatment scenarios could be completely different.[110]

Multiple vaccines are being developed, including mRNA-1273, Ad5-nCoV, INO-4800, LV-SMENP-DC and pathogen-specific artificial antigen-presenting cell (aAPC).[111] A general concern with these efforts is the extent, rate and impact of CoV mutations (e.g. of surface spike proteins) likely to occur over time after prolonged CNS infection and thus the efficacy and durability of potential protection.[87] Several mutations have been reported recently for SARS-CoV-2 in Non-Structural Protein 6 (NSP6), Open Reading Frame 10 (ORF10) adjacent region and S1 domain,[112,113] as well as changed binding capacity with ACE2.[114] One specific vaccine is unlikely to be enough, even if it is highly effective in a segment of the population. All vaccines also carry the potential for adverse effects, which means we have to be cautious in multiple ways.

To sum up, attention should be focused on the relationship between COVID-19 and neurodegenerative diseases for two main reasons: (1) SARS-CoV-2 infection may cause neuroinflammation which, in turn, may hasten the progression of some neurodegenerative disorders; (2) as neurodegenerative patients cannot take care of themselves, the quarantine itself may have negative consequences for the cognition and function, potentially contributing to their vulnerability to SARS-CoV-2 infection with exposure.

## Conclusion and Future Perspectives

Although the long-term consequences of SARS-CoV-2 infection in the brain are unclear, there is enough evidence in the literature to suggest that infection may exacerbate some CNS disorders. Resistance and resilience to any viral infection relies on a balanced immune response, which can be dysfunctional with aging and neurological diseases or their antecedents.[115] Given the prominence of respiratory symptoms, there is a risk for doctors to overlook consequences for the brain pathology. As we anticipate recurrent flare of infection, neurological care - more than ever - needs to be delivered using multi-disciplinary teams including neurorehabilitation, nutritional and social care professionals.[115,116] Characterising and understanding long-term outcomes of SARS-CoV-2 infections and other deleterious impacts now is a priority to be pursued through partnerships between academia and the NHS.

## Acknowledgements

PMM acknowledges generous personal and research support from the Edmond J Safra Foundation and Lily Safra and an NIHR Senior Investigator Award. This work also is supported by the UK Dementia Research Institute, which received its funding from UK DRI Ltd., funded by the UK Medical Research Council, Alzheimer's Society and Alzheimer's Research UK. Infrastructure was supported by the National Institute for Health Research (NIHR) Biomedical Research Centre (BRC). We also acknowledge the support from Getbodysmart website for a kind permission to reproduce their figure.

**Authors' contributions:** JT and SP wrote the primary draft. JT drawed all the figures, except for the one from Getbodysmart website. PMM and SG revised and edited the primary draft. All authors read and approved the final manuscript.